\documentclass[letterpaper,journal]{IEEEtran}
\usepackage{amsmath,amsfonts}
\usepackage{algorithmic}
\usepackage{algorithm}
\usepackage{array}
\usepackage[caption=false,font=normalsize,labelfont=sf,textfont=sf]{subfig}
\usepackage{textcomp}
\usepackage{stfloats}
\usepackage{url}
\usepackage{verbatim}
\usepackage{graphicx}
\usepackage{cite}
\usepackage[numbers]{natbib}
\usepackage{tabularx} % in preamble
\usepackage{color,soul}

\makeatletter
\def\bstctlcite#1{\@bsphack
  \@for\@citeb:=#1\do{%
    \edef\@citeb{\expandafter\@firstofone\@citeb}%
    \if@filesw\immediate\write\@auxout{\string\citation{\@citeb}}\fi
    \@ifundefined{bstctl@\@citeb}%
      {\global\@namedef{bstctl@\@citeb}{}}%
      {}%
  }%
  \@esphack}
\makeatother

\begin{document}

\title{DRL-Driven Edge-Aware Utility Optimization for Multi-Slice 6G Networks}

\author{Khaled M. Naguib,
Soumaya Cherkaoui, \IEEEmembership{Senior Member, IEEE}, Mahmoud M. Elmessalawy, \IEEEmembership{Member, IEEE}, Ahmed M. Abd El-Haleem, \IEEEmembership{Member, IEEE} and Ibrahim I. Ibrahim
\thanks{K. M. Naguib is with the CCAS Department, School of Engineering, New giza University (NGU), Cairo, Egypt, and with the Department of Computer and Software Engineering, Polytechnique Montreal, Montreal, QC H3T 1J4, Canada (e-mail: khaled.naguib@ngu.edu.eg).}
\thanks{S. Cherkaoui is with the Department of Computer and Software Engineering, Polytechnique Montreal, Montreal, QC H3T 1J4, Canada (e-mail: soumaya.cherkaoui@polymtl.ca).}
\thanks{M. M. Elmessalawy, A. M. Abd El-Haleem and I. I. Ibrahim are with the Department of Electronics and Communications, Faculty of Engineering, Helwan University, Cairo, Egypt (e-mail: melmesalawy@h-eng.helwan.edu.eg; ahmed\_abdelkhaliq@h-eng.helwan.edu.eg; iiibrahim1953@gmail.com).}

}

% The paper headers
\markboth{IEEE NETWORKING LETTERS,~Vol.~xx, No.~x, October~2025}%
{Shell \MakeLowercase{\textit{et al.}}: A Sample Article Using IEEEtran.cls for IEEE Journals}

\maketitle
\enlargethispage{15pt}

\IEEEpubid{\raisebox{-1.5\baselineskip}[0pt][0pt]{%
  \makebox[\columnwidth]{\hfill 0000--0000/00\textdollar00.00~\copyright~2021 IEEE\hfill}%
}}

\IEEEpubidadjcol

\begin{abstract}
Virtual Reality (VR) services delivered over 6G networks demand ultra-low latency and high bandwidth to ensure seamless user experiences. This paper presents an intelligent resource allocation and edge caching framework for 6G O-RAN networks, leveraging Deep Q-Network (DQN) learning for optimizing edge caching and dynamic resource provisioning across multiple network slices within an O-RAN-compliant architecture.By incorporating DRL agents into the network control plane, the proposed system enables proactive and adaptive content distribution as well as real-time computational resource allocation that meets the quality-of-service demands of eMBB, URLLC, and especially the emerging MBRLLC slices essential for VR. Simulation results demonstrate that the DQN-based framework consistently outperforms traditional methods in reducing latency and improving throughput, leading to more reliable and responsive support for immersive VR applications in 6G environments. 
\end{abstract}


\section{Introduction}

The rapid proliferation of immersive technologies—notably Virtual Reality (VR) and Augmented Reality (AR)—has intensified the demand for ultra-reliable, low-latency, and high-throughput communication systems. To address these stringent requirements, fifth-generation (5G) wireless networks and the forthcoming 6G paradigm introduce highly adaptive and intelligent architectures such as the Open Radio Access Network (O-RAN). O-RAN fundamentally restructures the radio access ecosystem by decoupling hardware from software components and enabling programmable, flexible management through RAN Intelligent Controllers (RICs). Specifically, the architecture distinguishes between the non-real-time RIC (nonRT-RIC), which coordinates centralized, policy-driven decision-making, and the near-real-time RIC (nRT-RIC), which supports low-latency, adaptive control. Both layers leverage advanced AI/ML-powered xApps and rApps to orchestrate network behavior dynamically in response to evolving service demands, thereby meeting the sophisticated performance goals of next-generation wireless networks (xApps and rApps) \cite{10433640}.

Simultaneously, edge computing and edge caching have emerged as prime facilitators for meeting the stringent latency and bandwidth requirements of today's services. By caching and processing content close to the users at the edge of the network, they are able to significantly reduce backhaul bottlenecks, improve responsiveness, and optimize overall Quality of Service (QoS), which is particularly critical for VR applications requiring low latency and high data rates. To address heterogeneous service requirements, network slicing allocates dedicated virtual resources for service types such as enhanced Mobile Broadband (eMBB), Ultra-Reliable Low-Latency Communications (URLLC), and newly introduced Mobile Broadband Reliable Low-Latency Communications (MBRLLC). Each slice supports a distinct performance profile: eMBB prioritizes high throughput, URLLC focuses on latency and reliability, while MBRLLC combines both aspects, making it suitable for demanding VR applications \cite{10702574}.

This work presents a centralized O-RAN architecture seamlessly integrated with edge computing and intelligent caching to support latency-sensitive, high-throughput services, with a focus on VR applications. The proposed framework advances conventional O-RAN by strategically deploying edge nodes at or near the Distributed Units (O-DUs), where they undertake edge computation, analytics, and content caching. These enhancements substantially minimize queuing and transmission delays, boost QoS metrics across all network slices, and enable fine-grained, slice-specific optimization using utility functions that capture the trade-offs between delay and throughput.

Recent research has examined the programmable and modular capabilities of O-RAN to achieve advanced resource management and network optimization. The authors in \cite{9999295} give an in-depth analysis of how network slicing in O-RAN architectures can be optimized concurrently for URLLC services with the assistance of federated deep reinforcement learning, specifically by controlling severe latency and reliability requirements. Expanding on this, the work the work in \cite{10921752} leverages quantum annealing to optimize O-RAN slicing for URLLC and eMBB services. Similarly, \cite{101145} showcases AI-driven xApps for resource management, emphasizing the role of centralized orchestration in optimizing multi-slice network environments.

The integration of edge computing with RAN architectures has garnered substantial attention in recent research . In \cite{10530215}, the authors present edge-native service orchestration for 5G, leveraging distributed micro data centers to enhance computational offloading and improve overall user experience. In a related direction,  \cite{10078092} examines the benefits of joint edge resource slicing, demonstrating how it can significantly reduce end-to-end latency. While these advancements demonstrably improve network responsiveness and efficiency, they do not directly address multi-slice optimization or the combination of these strategies within a control plane compliant with O-RAN.

Recent studies have shown progress in network slicing optimization. In \cite{70002}, a utility-based approach is presented for dynamically allocating resources between eMBB and URLLC slices in accordance with service-level agreements. The introduction of MBRLLC slicing in \cite{10765787}, which unites the strengths of eMBB and URLLC, is found to be particularly advantageous for applications demanding both high throughput and reliability, such as VR. Nonetheless, these approaches do not consider the joint influence of edge computing and caching on the performance of network slices. In \cite{10884912}, the authors examine VR streaming over 6G and find that edge caching of popular VR content reduces playback delays and can benefit user experience. Their delay-sensitive VR delivery framework is designed for MEC environments and mainly addresses responsiveness. Integration with dynamic slice management or centralized O-RAN control is not explored within the scope of this work.

The literature thus reveals two critical research gaps: (i) no comprehensive framework exists to combine edge computing and caching within the O-RAN architecture to maximize slice-specific benefits, and (ii) there are limited tailored mechanisms supporting MBRLLC slices for VR applications. Our model seeks to overcome these shortcomings by introducing a utility-driven optimization strategy, which factors in content popularity, cache status, and network conditions, and is realized within a centralized O-RAN system augmented by edge-enabled O-DUs.

\section{System Model and Problem Formulation}

The proposed system model employs a centralized O-RAN architecture enhanced with integrated edge computing and edge caching. This design focuses on reducing latency, enabling computation offloading, and alleviating backhaul traffic. Resource allocation spans across eMBB, URLLC, and MBRLLC network slices, with utility functions driving slice-specific optimization and real-time adaptation to varying user needs and channel conditions. Resource allocation is performed over eMBB, URLLC, and MBRLLC slices with utility functions guiding slice-specific optimization, and adaptation to user demands and channel conditions. Within this framework, VR users are primarily assigned to MBRLLC slices, reflecting their requirements for high data rates and ultra-low latency. The architecture deploys edge nodes equipped with computational and storage resources at or near the O-DUs, effectively minimizing delays for VR applications.The edge nodes facilitate real-time analytics, local processing, and content caching, thus offloading the centralized nRT-RIC and minimizing core network congestion. 

\begin{figure}
\centering
\includegraphics[width=\columnwidth]{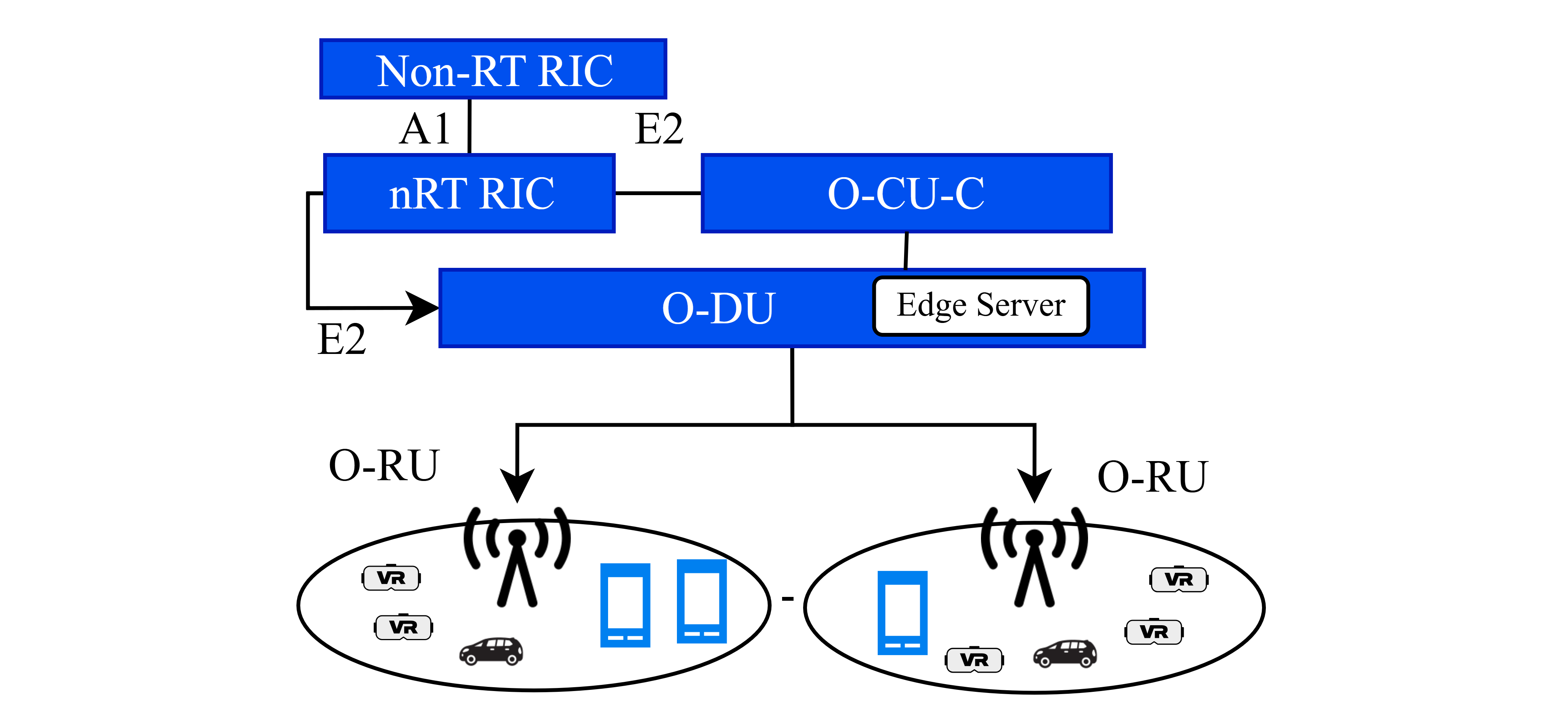} % Adjust width as needed
\caption{System Model}
\label{System model}
\end{figure}

As illustrated in in Fig. \ref{System model}, this scenario encompasses several O-RAN–compliant components, including O-DUs, O-RUs, and O-CUs, collectively functioning as base stations (BSs), indexed by $m \in M$. All the BSs are managed by a centralized NonRT-RIC and interact with a proprietary xAPP in the nRT-RIC for precise control. Edge computing modules co-located with or near each O-DU  improve this architecture by providing essential local computation and storage capabilities, critical for latency-sensitive and high-throughput applications. To fully exploit edge computing, intelligent caching mechanisms are employed, whereby popular VR content for MBRLLC slices is proactively stored and managed through a cache index table and access popularity statistics. These statistics are collaboratively updated among edge nodes, optimizing caching efficiency across the network.

Each slice $s \in \mathcal{S}$ is managed with consideration for content popularity and specific caching needs, with users for each slice represented by $N_s$. To further reduce latency and backhaul load, each base station $m$ is equipped with a finite cache capacity $C_m$. User content requests are denoted by $F_u$. The cache hit indicator, $H_{u,m}^{s}$ is defined as 1 if the requested content by user $u$  in slice $s$ is cached at BS $m$, and 0 otherwise.

Thus, the transmission delay is updated to account for cache hits:
\begin{equation}
    d_{u,m}^{s,trans} = 
    \begin{cases}
        \frac{L_u}{R_{u,m}^{s}} & \text{if } H_{u,m}^{s} = 0 \\
        \frac{\delta L_u}{R_{u,m}^{s}} & \text{if } H_{u,m}^{s} = 1
    \end{cases}
\end{equation}

where $0 < \delta < 1$ is a compression factor that is the lighter load from cached, pre-processed, or fragmented content. This improvement affects transmission delay and queuing delay, and therefore the total delay:

\begin{equation}
d_{u,m}^{s} = d_{u,m}^{s,trans} + \frac{1}{\frac{R_{u,m}^{s}}{L_u} - \eta_u}
\end{equation}

where $\eta_u$ is the packet arrival rate of user u. To model the impact of edge caching on service quality over different network slices, we suggest a shared utility function $U_{u,m}^{s}$ representing slice-dependent performance demands. The utility changes according to either throughput, delay, or both, depending on the service type, and comprises the value of edge caching via a cache gain term. The function can be defined as:

{\small
\begin{equation}
U_{u,m}^{s} =
\begin{cases}
\displaystyle \frac{2}{\pi}\tan^{-1}\left(R_{u,m}^{s} \right), & \text{if } s = \text{eMBB} \\[6pt]
\displaystyle 1 - \frac{2}{\pi}\tan^{-1}\left(d_{u,m}^{s} \left(1- \kappa H_{u,m}^{s}\right)\right), & \text{if } s = \text{URLLC} \\[6pt]
\displaystyle \frac{1}{2} \left[1 - \frac{2}{\pi} \tan^{-1}\left(d_{u,m}^{s} \left(1- \kappa H_{u,m}^{s}\right)\right) \right. \\[2pt]
\quad\quad\quad\quad + \left. \frac{2}{\pi} \tan^{-1}\left(R_{u,m}^{s} \right)\right], & \text{if } s = \text{MBRLLC}
\end{cases}
\end{equation}
}

Here, $R_{u,m}^{s}$ and $d_{u,m}^{s}$ are the achievable rate and experienced delay for user $u$ in base station $m$ of slice $s$, respectively. $\kappa$ is the weight assigned to the advantage of the cache from the delay perspective. This account provides a holistic, slice-aware explanation of how edge caching enhances user-perceived usefulness and supports differentiated Service Level Agreements (SLAs).

The problem is described as to maximize the aggregated slice-aware utility across all slices which is subjected to caching constraints and SLA requirements. Thus, the problem can be formulated as following:

\begin{align}
\max_{a} \quad & \sum_{s \in \mathcal{S}} \sum_{u \in N_s} \sum_{m \in \mathcal{M}} U_{u,m}^{s} \\
\text{s.t.} \quad 
& d_{u,m}^{s} \leq D_{th,m}, \quad \forall u, m, s \tag{i} \\
& R_{u,m}^{s} > R_{th,m}, \quad \forall u, m, s \tag{ii} \\
& \sum_{f \in F_u} S_{f,m} \cdot \text{size}(f) \leq C_m, \quad \forall m \in \mathcal{M} \tag{iii} \\
& H_{u,m}^{s} \in \{0, 1\}, \quad \forall u, m, s \tag{iv}
\end{align}

Here, $a$ represents the joint decision variables, including cache placement indicators $H_{u,m}^{s}$ and resource allocation parameters. In constraint (iii), $S_{f,m}$ is a binary variable indicating whether content $f$ is cached at base station $m$; this ensures the sum accounts only for unique cached contents, thereby avoiding duplicate counting of files requested by multiple users. Constraints (i) and (ii) enforce the per-user delay and rate requirements (SLAs) for each slice at every BS. Constraint (iii) maintains that the total size of cached content at each BS does not exceed its storage capacity, and (iv) restricts cache placement indicators to binary values. The vector $a$ collects all resource and caching decisions.
This formulation captures the interoperability between QoS-aware slicing and edge caching, thus fulfilling the differentiated requirements of various services.

\section{DQN-Based Resource Allocation Optimization}

To address the intricate problem of resource allocation and caching problem of the proposed O-RAN-based system model, we adopt a Deep Q-Network (DQN) solution. The goal is to maximize the overall slice-specific utility by simultaneously optimizing the transmission rate and caching choices, with user delay sensitivity, throughput demands, and limited edge storage. Classical optimization methods are also put to test in this environment's dynamic, high-dimensional nature—changes in user requirements, channel fluctuation, and heterogeneity of QoS requirements within the slices of a network. Reinforcement learning (RL), particularly DQN, gives the possibility of a scalable, adaptive scheme that can learn efficient policies by engaging with the environment.

The structure is established as an MDP, where a centralized agent (say, an xApp in the nRT-RIC) acts upon the O-RAN system at each decision interval. The state at time $s_t$
includes channel quality indicator encoding, cache status, user demand profiles, arrival rates, and real-time resource utilization at each base station. The agent learns an action $a_t$ comprising discrete decisions on resource allocation—e.g., assigning data rates and cache updates, which indicate whether a specific content item is stored at edge cache for user $u$ in slice $s$.

The reward signal $r_t$ guides the learning process and records the short-term utility benefit and penalty breaking for violating SLA limits. Specifically, it includes the sum of per-user utility values which capture slice-specific performance (delay and throughput) and weigh in the benefit of cache hits, deducting penalties for cache overflow and high delay. The DQN agent attempts to maximize the expected cumulative discounted reward by learning the optimum action-value function $Q(s,a)$, learned through a deep neural network. In each step, the network approximates Q-values for all actions from the current state, and the agent takes actions according to an $\epsilon$-greedy policy to balance exploration and exploitation. Transitions $(s_t,a_t,r_t,s_{t+1}$ are stored in a replay buffer, and the network is trained on mini-batches of past experiences to stabilize learning and improve sample efficiency.

The training process continues until the Q-network converges on a plateau of the moving average of cumulative rewards. Once trained, the DQN agent can be deployed in real-time in the RAN Intelligent Controller to make adaptive, slice-aware control decisions. By incorporating this model within an xApp, the system gains the ability to dynamically allocate resources and cache strategies based on observed network conditions, without undergoing full-scale reoptimization at every instant. This enables significant improvements in throughput for eMBB, latency reduction for URLLC services, and also fairness of performance for MBRLLC users, especially in VR scenarios where high rates and low delay are critical. Furthermore, by providing compute and storage locally at the edge nodes, the system reduces the dependency on the central network and improves scalability.

Overall, the DQN framework presents an intelligent, flexible approach for orchestrating caching and resource allocation within a heterogeneous O-RAN environment supporting diverse QoS demands. It allows the system to autonomously develop efficient control policies that jointly optimize latency, bandwidth, caching effectiveness, and nuanced, slice-specific service requirements. The computational costs of the proposed DQN-based solution are explicitly considered to ensure its viability in O-RAN deployments. While offline training is carried out in the non-RT RIC with sufficient computational resources, the inference phase in the near-RT RIC involves only lightweight forward passes over the trained neural network. This allows real-time decision-making, as the inference overhead is low compared to the timescales of slice-level resource allocation. Consequently, the proposed method can be deployed in live O-RAN systems without violating the extremely tight latency constraints of 6G services.

\section{Results and Discussion}

The proposed DQN-based cache and resource allocation scheme is evaluated in a simulated 6G O-RAN network comprising 7 base stations (BSs). Each BS serves 42 mobile users distributed across a 150-meter coverage area. Users are assigned to one of three network slices; eMBB, URLLC, or MBRLLC, according to their service requirements. Mobility is modeled using the SUMO traffic simulator, where pedestrian users move at speeds between 1 and 3 meters per second. This mobility model is integrated with the NS-3 network simulator to generate realistic network and channel dynamics throughout the simulation. Each BS is provisioned with 10 MHz of bandwidth, partitioned into 50 resource block groups, and the maximum user transmission power is set at 23 dBm. Traffic patterns reflect the distinct characteristics of each network slice.  URLLC users generate 200 KB packets at a rate of 10 packets per second. In contrast, eMBB and MBRLLC users transmit 2 MB packets at 30 packets per second, mirroring their higher throughput requirements. The DQN agent monitors user state, channel conditions, and cache availability to determine optimal strategies for both resource allocation and content caching. The agent is trained to maximize a slice-dependent utility function that dynamically balances user throughput and latency, in accordance with the unique QoS demands of each slice. By employing prioritized experience replay and soft target update, the DQN achieves stable convergence and effectively adapts to dynamic network conditions caused by user mobility and diverse QoS requirements. This approach allows the DQN to outperform both static and fairness-driven baselines with real-time, mobility-aware decision-making that meets the distinct performance goals of each slice.

\begin{figure}
\centering
\includegraphics[width=\columnwidth]{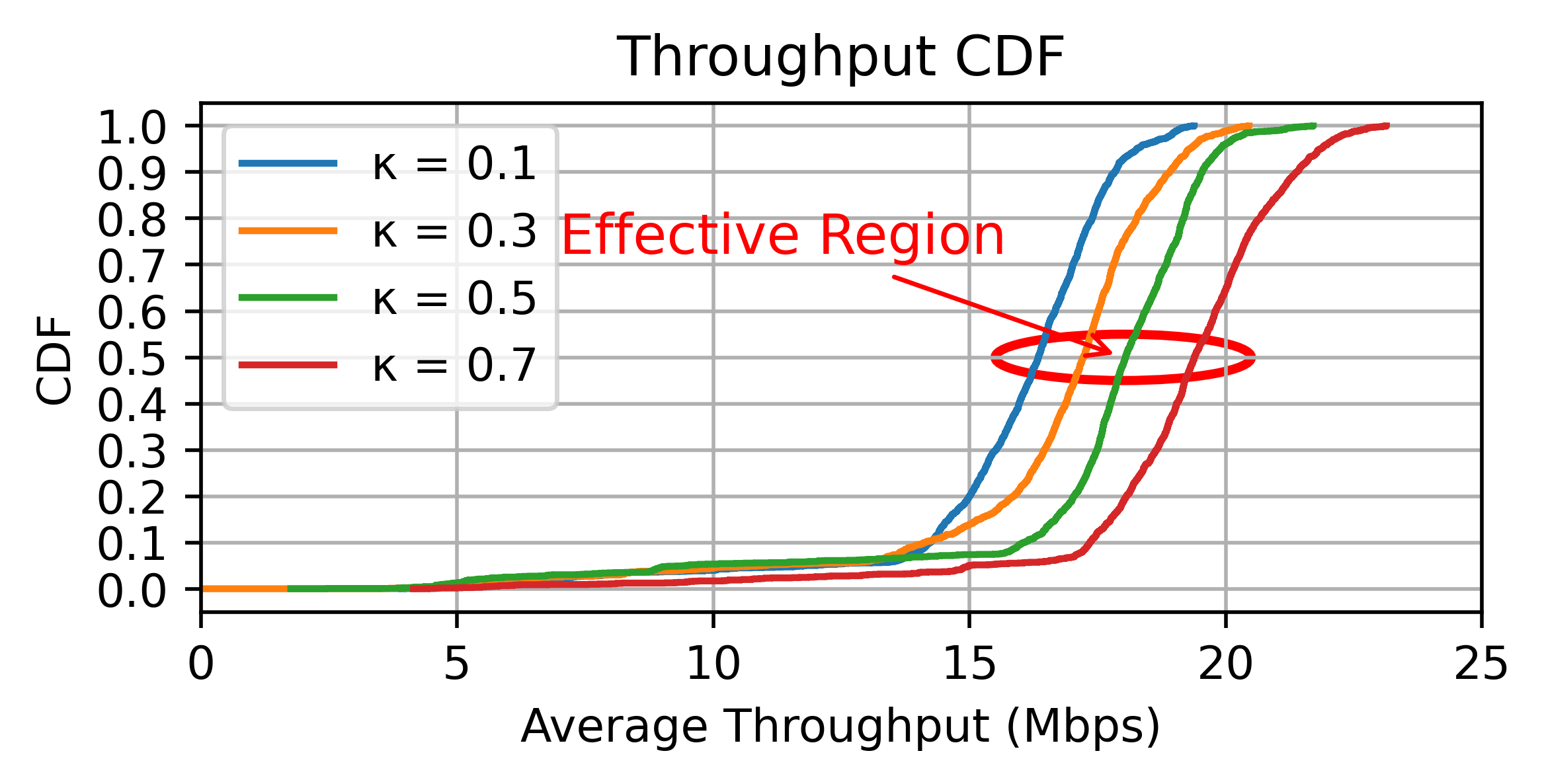} % Adjust width as needed
\caption{Effect of Cache Gain Weight ($\kappa$) on Throughput Distribution}
\label{Throughput CDF}
\end{figure}

Fig. \ref{Throughput CDF} show the Cumulative Distribution Function (CDF) of average user throughput for various values of the cache gain weight parameter $\kappa$, which is the delay reduction benefit contribution of edge caching to the utility function. Intuitively, $\kappa$  acts as a balancing parameter between cache benefits and effective radio resource allocation: low values under-emphasize the impact of caching, while excessively high values over-prioritize caching at the expense of radio resource usage. An increase in $\kappa$ causes a shift of the throughput distribution to the right, indicating improved throughput performance for more users. In our experiments, $\kappa$ = 0.7 always outperforms other setups in the area of effective performance at the 50th percentile, so that increasing the focus on caching gain is leading to more resource-aware decisions taken by the DQN agent. This because  $\kappa$ = 0.7 provides an appropriate trade-off, giving sufficient priority to caching to reduce latency through local content delivery, while allowing the DQN agent enough flexibility to optimize radio resource allocation effectively. This gain is particularly apparent for moderate channel users experience, where transmission delay and queuing delay are alleviated by the cache servers, allowing for more efficient reuse of RBs. The highlighted region shows that utility design cache contribution balancing is essential for enhancing overall spectral efficiency and user QoS.

\begin{figure}
\centering
\includegraphics[width=\columnwidth]{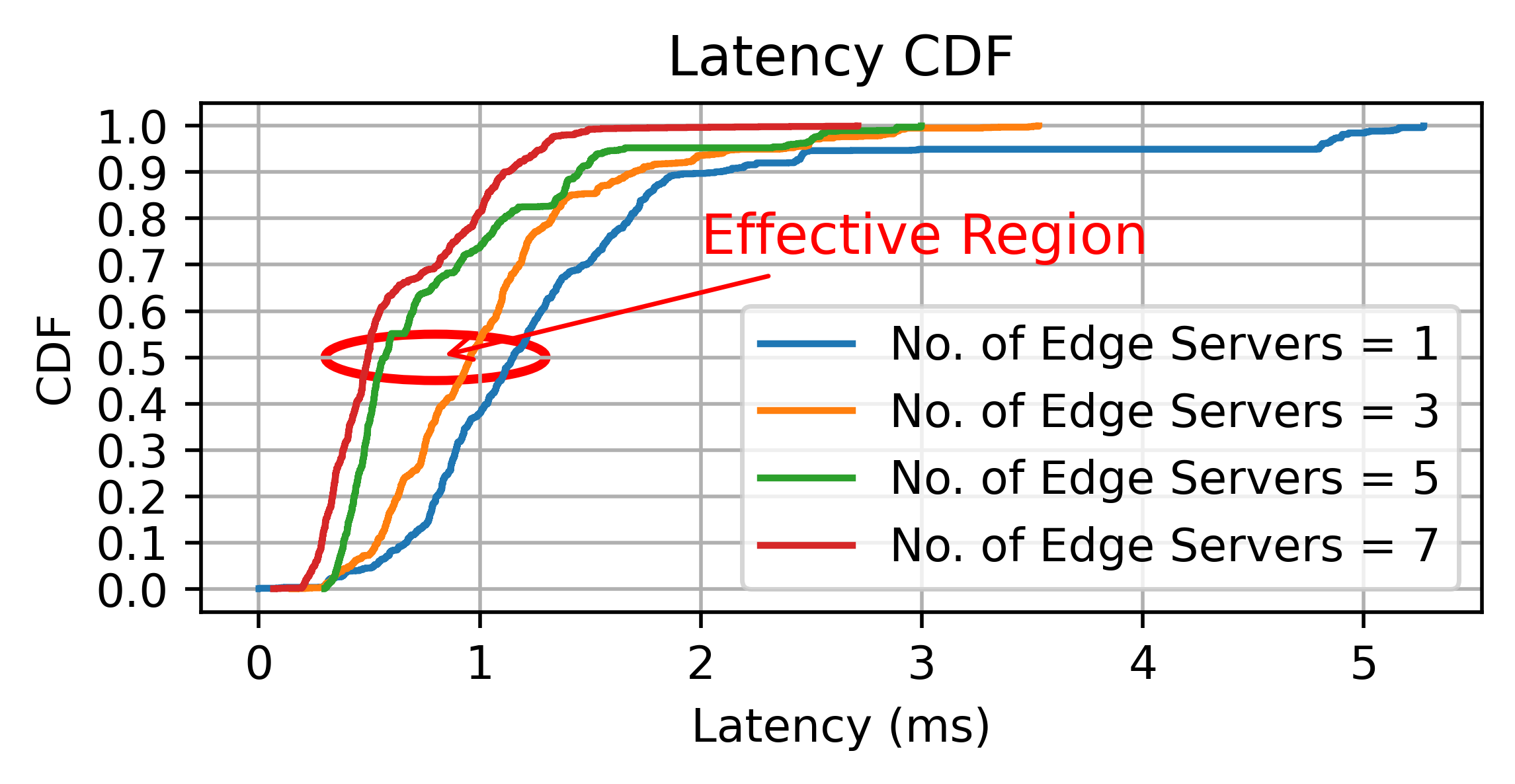} % Adjust width as needed
\caption{Impact of Edge Server Density on Latency Distribution}
\label{Latency CDF}
\end{figure}

Fig. \ref{Latency CDF} presents the CDF of the latency experienced by users with varying numbers of edge servers. The latency distribution significantly improves with the growing number of edge servers, i.e., from 1 to 7, since the CDF curves shift to the left. This is to say that there are increasingly more users who experience lower latencies, especially in the effective region near the median. The 7 edge server setting has the sharpest slope and lowest values of latency, illustrating that task offloading effectiveness and transmission latency reduce with higher availability of edge servers. The results demonstrate that the DQN-based resource allocation model effectively adapts to changes in edge infrastructure, optimizing performance for latency-sensitive traffic such as URLLC by enabling more responsive and delay-aware scheduling decisions.

\begin{table}[ht]
\centering
\caption{Fairness vs. number of edge caching servers. Jain’s Fairness Index (JFI) is computed as 
$JFI(X)=\frac{(\sum_{i=1}^{n} x_i)^2}{n \cdot \sum_{i=1}^{n} x_i^2}$, 
where $x_i$ are per-UE satisfaction scores within a slice.}
\label{tab:jfi_edge_servers}
\renewcommand{\arraystretch}{1.15}
\begin{tabularx}{\columnwidth}{c *{4}{>{\centering\arraybackslash}X}}
\hline
\textbf{\# Servers} & \textbf{JFI\textsubscript{eMBB}} & \textbf{JFI\textsubscript{URLLC}} & \textbf{JFI\textsubscript{MBRLLC}} \\
\hline
1 & $0.82 $ & $0.78 $ & $0.80$ \\
3 & $0.88$ & $0.83$ & $0.85 $ \\
5 & $0.91 $ & $0.86 $ & $0.88 $ \\
7 & $\mathbf{0.92} $ & $\mathbf{0.87} $ & $\mathbf{0.89} $  \\
\hline
\end{tabularx}
\vspace{2pt}

\end{table}

As shown in Table ~\ref{tab:jfi_edge_servers}, fairness improves as the number of edge caching servers increases. The trend is clear: higher caching density enhances overall fairness across all slices, as locally available hot content reduces backhaul congestion and alleviates contention arising from varying slice demands. For instance, the JFI for eMBB and MBRLLC rises from approximately 0.80 with one caching server to above 0.90 with seven servers, while URLLC fairness improves progressively in line with its strict latency and reliability requirements. Cross-slice JFI also increases, indicating a more equitable allocation of resources across service types. These results confirm that increased caching density promotes fairer performance under DQN-based control with slice-specific utility shaping.

\begin{table}[ht]
\centering
\caption{Impact of caching, DQN optimization, and slice-specific utility shaping on performance metrics.}
\label{tab:performance}
\fontsize{8}{9}\selectfont
\resizebox{\columnwidth}{!}{%

\begin{tabular}{lccc}
\hline
\textbf{Configuration} & \textbf{Latency (ms)} & \textbf{Throughput (Mbps)} & \textbf{SLA Satisfaction (\%)} \\ \hline
Baseline (no caching, no DQN, no shaping) & 6.1  & 10.2 & 71.6 \\
+ Caching                                 & 4.3  & 15.4 & 78.3 \\
+ DQN                                     & 2.8  & 17.2 & 83.1 \\
+ Utility shaping (latency-focused)       & 0.83  & 12.1 & 86.7 \\
+ Utility shaping (throughput-focused)    & 2.1  & 18.4 & 88.9 \\
Full model & 0.72 & 19.3 & 92.8 \\ \hline
\end{tabular}
}
\end{table}

Table~\ref{tab:performance} summarizes the individual contributions of each component. Caching improves end-to-end performance by reducing latency through higher cache-hit ratios, while DQN optimization enhances throughput by adapting to dynamic network conditions. Slice-specific utility shaping further optimizes resource allocation according to the requirements of each slice: shaping for latency targets reduces delay for delay-sensitive slices, whereas shaping for throughput targets increases capacity for bandwidth-demanding slices. The complete model achieves an optimal balance among all measures, highlighting the complementary benefits of these mechanisms.

\section{Conclusion}

This works presents an intelligent resource allocation and edge caching framework for 6G O-RAN networks, leveraging DQN to address the diverse and dynamic requirements of multiple network slices. By integrating edge computing and proactive caching capabilities at O-RAN DUs, our approach enables adaptive, slice-aware decisions that account for real-time mobility, content popularity, and stringent service-level agreements. Through detailed simulations, we demonstrated that the proposed DQN-based scheme outperforms conventional methods in supporting both low-latency services (such as VR over MBRLLC and URLLC slices) and high-throughput applications. The model dynamically balances delay and data rate optimization, providing robust quality of experience across variable network conditions and user mobility patterns. These results suggest that DRL-powered control, combined with architectural flexibility at the edge, offers an effective pathway for next-generation RANs to meet the heterogeneous demands of emerging immersive applications and differentiated network services. The proposed xApp can be deployed in ORANSlice by interfacing with the near-RT RIC via standard E2 and O1 interfaces. This enables real-time collection of slice-specific telemetry, DQN-based resource optimization, and enforcement of slice-aware allocation decisions, allowing evaluation of latency, throughput, and fairness under realistic network conditions. This demonstrates the practical feasibility and effectiveness of the approach in operational O-RAN environments.

\section*{ACKNOWLEDGEMENT}
\label{sec:acknowledgement}
The authors would like to thank the Natural Sciences and Engineering Research Council of Canada, and the The Fonds de recherche du Québec – Nature et technologies (FRQNT), for the financial support of this research.

% Generated by IEEEtran.bst, version: 1.14 (2015/08/26)

\end{document}